\documentclass[10pt,a4paper,onecolumn]{article}

\usepackage{graphicx}
\usepackage{dcolumn}
\usepackage{bm}
\usepackage{hyperref}
\usepackage{url}
\usepackage{color}
\usepackage{authblk}
\usepackage{amsmath}
\usepackage{amsfonts}
\usepackage{amssymb}
\usepackage{placeins}
\usepackage{helvet}

\usepackage{setspace,caption}
\usepackage[]{fullpage}
\usepackage{amsmath}
\usepackage[left]{lineno}

\newcommand{\pcto}{PbCuTe$_{2}$O$_{6}$}
 
\title{Evidence for a three-dimensional quantum spin liquid in \pcto}

\author[1,*]{\small S. Chillal}
\author[2]{\small Y. Iqbal}
\author[3]{\small H. O. Jeschke}
\author[4,5]{\small J. A. Rodriguez-Rivera}
\author[6]{\small R. Bewley}
\author[6]{\small P. Manuel}
\author[6]{\small D. Khalyavin}
\author[7]{\small P. Steffens}
\author[8]{\small R. Thomale}
\author[1]{\small A. T. M. N. Islam}
\author[1,9]{\small J. Reuther}
\author[1,10]{\small B. Lake}

\affil[1]{\footnotesize Helmholtz-Zentrum Berlin f\"ur Materialien und Energie, Hahn-Meitner Platz 1, 14109 Berlin, Germany}
\affil[2]{\footnotesize Department of Physics, Indian Institute of Technology Madras, Chennai 600036, India}
\affil[3]{\footnotesize Research Institute for Interdisciplinary Science, Okayama University, 3-1-1 Tsushima-naka, Kita-ku, Okayama 700-8530, Japan}
\affil[4]{\footnotesize NIST Center for Neutron Research, National Institute of Standards and Technology, 20899 Gaithersburg, USA}
\affil[5]{\footnotesize Department of Materials Science, University of Maryland, College Park, 20742 Maryland, USA}
\affil[6]{\footnotesize ISIS Facility, STFC Rutherford Appleton Laboratory, Oxfordshire OX11 0QX, UK}
\affil[7]{\footnotesize Institut Laue-Langevin, 71 Avenue des Martyrs, 38042 Grenoble Cedex 9, France}
\affil[8]{\footnotesize Institute for Theoretical Physics and Astrophysics,
Julius-Maximilian's University of W\"urzburg, Am Hubland, D-97074 W\"urzburg, Germany}
\affil[9]{\footnotesize Dahlem Center for Complex Quantum Systems and Fachbereich Physik, Freie Universit\"at Berlin, 14195 Berlin, Germany}
\affil[10]{\footnotesize Institut f\"ur Festk\"orperphysik, Technische Universit\"at Berlin, Hardenbergstr. 36, 10623 Berlin, Germany}
\affil[*]{\footnotesize Correspondence: shravani.chillal@helmholtz-berlin.de}

\date{\today}
\begin{document}

\maketitle
\clearpage
{\bf The quantum spin liquid (QSL) is a highly entangled magnetic state characterized by the absence of static magnetism in its ground state. Instead, the spins fluctuate in a highly correlated way down to the lowest temperatures. The QSL is very rare and is confined to a few specific cases where the interactions between the magnetic ions cannot be simultaneously satisfied (known as frustration). Lattices with magnetic ions in triangular or tetrahedral arrangements which interact via isotropic antiferromagnetic interactions can generate such a frustration. Three-dimensional isotropic spin liquids have mostly been sought in materials where the magnetic ions form pyrochlore or hyperkagome lattices. Here we present a three-dimensional lattice called the hyper-hyperkagome that enables spin liquid behaviour and manifests in the compound \pcto. Using a combination of experiment and theory we show that this system exhibits signs of being a quantum spin liquid with no detectable static magnetism together with the presence of diffuse continua in the magnetic spectrum suggestive of fractional spinon excitations.}

Fractionalization is one of the most fascinating phenomena in modern condensed matter physics. In the context of spin systems, a paradigmatic example of fractionalization occurs for the one-dimensional (1D) magnet formed from half-integer spin magnetic ions coupled by isotropic antiferromagnetic interactions. In this system, the spins cannot order at any temperature $T>0$~{\rm K}, and they exhibit exotic excitation spectra. Even though the spin operators in the Hamiltonian can only flip the spins by an integer unit of Planck's constant, the actual excitations carry spin $S=\frac{1}{2}$ and are called spinons. Since any experimental technique can only change the angular momentum by an integer unit, spinons cannot be created singly but only in multiple pairs, leading to the characteristic multi-particle excitation continua observed in neutron scattering experiments~\cite{lake2005,Lake2013,Mourigal2013}. 
The concept of fractionalization and spinons can be extended to two and higher dimensions. 
In this case, spinons occur as excitations of quantum spin liquids - which have no static magnetism even at $T=0$~{\rm K}, i.e. they require an environment of strong quantum fluctuations as realized by magnets with small spin magnitudes and frustrated interactions~\cite{Bal10}. Examples of two-dimensional (2D) quantum spin-liquid candidates are the kagome materials Herbertsmithite (ZnCu$_3$(OH)$_{6}$Cl$_{2}$)~\cite{Han2012} and Ca$_{10}$Cr$_{7}$O$_{28}$~\cite{Balz2016,Balz2017} where the $S=\frac{1}{2}$ moments form a highly-frustrated network of corner-sharing triangles and the excitations form broad, diffuse and dispersionless ring-like features suggestive of multi-spinon continua. This behaviour is in stark contrast to the spin-wave excitations characterized by spin $S=1$ which are observed as sharp modes in magnets with conventional static magnetic order. 

Fractionalization can also be achieved in three-dimensional (3D) spin systems. The pyrochlore lattice which consists of a three-dimensional network of corner-sharing tetrahedra can support a number of spin liquid states~\cite{Canals1998,Moessner1998,Benton2016,Savary2017,Iqbal2017,Iqbal2019}. 
The most famous of these is the spin ice state which arises from the combination of strong local Ising anisotropy and ferromagnetic interactions as found for Dy$_2$Ti$_2$O$_7$ and Ho$_2$Ti$_2$O$_7$~\cite{fen09,mor09}. Spin ice is in fact a classical spin liquid characterized by a macroscopic ground state degeneracy and fractional monopole excitations. It is worth emphasizing that classical spin liquids are distinctly different from quantum spin liquids and only exist for large local moments and/or very strong local anisotropies. In the case of spin ice, a reduction in the strength of the anisotropy induces quantum tunneling between these classically degenerate states. This lifts the degeneracy giving rise to quantum spin ice where the spins are no longer static but fluctuate coherently in the ground state~\cite{Gingras2014,Benton2012}. In contrast to the advanced exploration of the pyrochlore spin liquids, current understanding of 3D networks of corner-sharing triangles as exemplified by the hyperkagome lattice are much less developed, although they are also expected to support spin liquid behavior in the presence of isotropic antiferromagnetic interactions~\cite{Hopkinson2007,Zhou2008,Bergholtz2010}. There are two physical realizations of the hyperkagome lattice, Gd$_3$Ga$_5$O$_{12}$ which has large ($S=\frac{7}{2}$) magnetic ions and is proximate to a classical spin liquid~\cite{Schiffer1995,Quilliam2013}, while Na$_4$Ir$_3$O$_8$ has quantum ($S=\frac{1}{2}$) ions and is proximate to a quantum spin liquid~\cite{Dally2014}.

\pcto is a three-dimensional magnet consisting of $S=\frac{1}{2}$ Cu$^{2+}$ magnetic moments coupled by isotropic antiferromagnetic interactions  into a three-dimensional network of corner-sharing triangles. Synthesis and measurements of powder samples of this compound have been previously reported~\cite{Koteswararao2014,Khuntia2016}. DC magnetic susceptibility yields a Curie-Weiss temperature of $\theta_{\rm CW}=-22$~{\rm K} revealing predominantly antiferromagnetic exchange interactions~\cite{Koteswararao2014}. Although heat capacity data on a powder show broad features at temperatures around $T\sim1$~{\rm K}~\cite{Koteswararao2014}, they do not reveal a sharp $\lambda$-type anomaly and there is no phase transition to long-range magnetic order. Muon spin relaxation measurements also confirm the absence of any static magnetism down to $0.02$~{\rm K}~\cite{Khuntia2016}. They do however, reveal enhanced magnetic correlations at low temperatures which onset below $T<1$~{\rm K} implying the presence of persistent spin dynamics in the ground state as is expected of a quantum spin liquid.

In the following, we provide strong experimental and theoretical indications for the quantum spin liquid state in \pcto. We perform neutron diffraction and inelastic neutron scattering measurements and show that the absence of long-range magnetic order in the powder sample of this compound is accompanied by diffuse spheres of dispersionless inelastic scattering consistent with a multi-particle continuum of spinons. We also determine the exchange interactions using density functional theory and establish that a three dimensional frustrated motif called the hyper-hyperkagome is responsible for this behaviour. Finally, we reproduce the observed ground state and dynamics using pseudo-fermion functional renormalization group calculations and show that this Hamiltonian generates the behaviours associated with a quantum spin liquid. The strength of our work lies at the excellent agreement between the experiment and theory that has not been observed until now for quantum spin liquids arising in such complex three dimensional systems.

\section*{Results}
\subsection*{Magnetic interactions in \pcto}
The positions of the magnetic Cu$^{2+}$ ions in \pcto are shown in Fig.~\ref{Figure:1}a where the green and red bonds represent the 1$^{\rm st}$ and 2$^{\rm nd}$ nearest neighbour interactions $J_{1}$ and $J_{2}$, respectively. All  Cu$^{2+}$ ions are crystallographically equivalent. On its own, $J_{1}$ couples the Cu$^{2+}$ moments into isolated triangles, while $J_{2}$ forms a three-dimensional network of corner-sharing triangles known as the {\it hyperkagome} lattice. Further neighbour interactions are also shown in Fig.~\ref{Figure:1}b where, the 3$^{\rm rd}$ neighbour interaction $J_{3}$, forms isolated chains running parallel to the crystalline {\bf a}, {\bf b}, {\bf c} axes and the 4$^{\rm th}$ neighbour interaction $J_{4}$, is responsible for chains parallel to the body diagonals. The complete Hamiltonian is
\begin{equation}
\mathcal{H}=\sum_{k=1,2,3,4(i<j)}J_{k} \mathbf{S}_{i}\cdot\mathbf{S}_{j},
\label{eq:1}
\end{equation}
where the interactions are assumed to be spin-isotropic, thus allowing the spins to point in any direction. This is justified because, as a light transition metal ion with only one hole in the $3d$ shell, the orbital moment of Cu$^{2+}$ is quenched by the strong square-planar crystal field due to the surrounding O$^{2-}$ ions. DC susceptibility confirms that the interactions are spin-isotropic, since it is almost independent of the direction of the applied magnetic field (Supplementary Note 1).

\subsection*{The ground state}
\begin{figure}
\centering
\includegraphics[width=1 \textwidth]{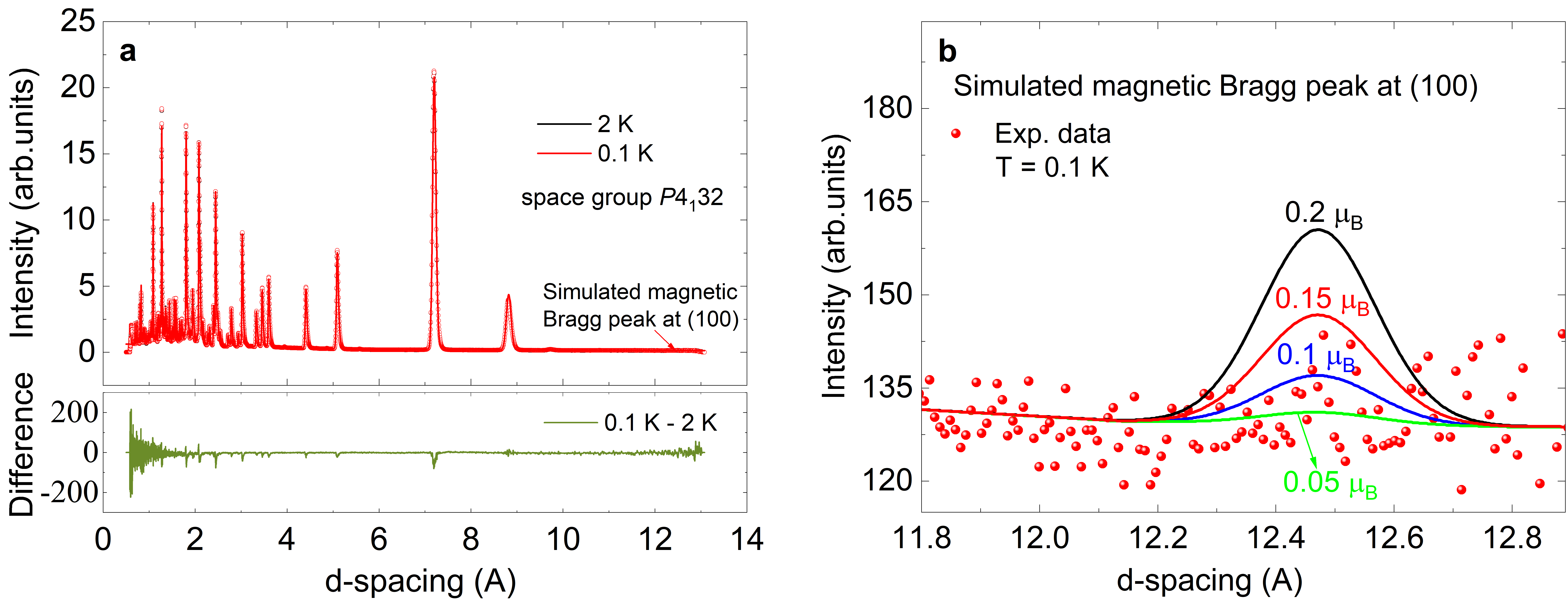}
\caption{\sffamily  \textbf{Powder neutron diffraction patterns of \pcto.} \textbf{a} Measured at $0.1$~{\rm{K}} and $2$~{\rm{K}} using the WISH high-flux diffractometer and plotted as a function of d-spacing. Both patterns are refined in the established cubic space group, $P\,4_{1}32$~\cite{Wulff1997} where the lattice constant at $0.1$~{\rm{K}} is $12.4454(3)$~\AA. The difference between the patterns at these two temperatures is plotted below in green and clearly shows that no magnetic Bragg peaks appear at the base temperature. \textbf{b} Assuming a magnetic structure compatible with the long range magnetic order found in SrCuTe$_{2}$O$_{6}$~\cite{Chillal2020}, a magnetic Bragg peak is estimated at the $(1,0,0)$ reflection ($12.4454$~\AA~ in d-spacing). The expected Bragg peak amplitude is shown for different values of ordered moment per Cu$^{2+}$ ion by the curves. Clearly, if present, the maximum ordered moment can be no greater than $0.05~\mu_{B}$.}
\label{Figure:2}
\end{figure}

In agreement with previous specific heat~\cite{Koteswararao2014} and muon spin relaxation~\cite{Khuntia2016} (Supplementary Note 2) measurements we did not find any evidence for long-range magnetic order or static magnetism in powder samples of \pcto down to $20$~mK. We also performed neutron diffraction which directly measures the spatial Fourier transform of the spin-spin correlation function and would show resolution-limited magnetic Bragg peaks in the case of long-range magnetic order.  Figure~\ref{Figure:2}a shows the neutron powder diffraction patterns of \pcto measured at temperatures $T=2$~{\rm K} and $0.1$~{\rm K} above and below $T\sim1$~{\rm K} respectively where features are observed in the specific heat and muon spin relaxation. Both patterns can be described entirely by considering only the known crystal structure of \pcto~\cite{Wulff1997}. The absence of any additional Bragg peaks that could correspond to long-range magnetic order is further revealed by taking the difference between the diffraction patterns at these two temperatures as shown by the lower green curve. To establish an upper limit on the maximum size of any possible static ordered moment, several magnetic structures were simulated and compared to the data. Figure~\ref{Figure:2}b shows a modeled magnetic Bragg peak compatible with the magnetic structure of the iso-structural compound SrCuTe$_{2}$O$_{6}$ for different sizes of the ordered moment~\cite{Chillal2020}. The ordered moment if present must be smaller than $\approx0.05~\mu_{B}$/Cu$^{2+}$ which is much less than the total spin moment of the Cu$^{2+}$ ion of $1~\mu_{B}$ indicating that static magnetism is suppressed. It should be mentioned that single crystals of \pcto do show signs of a phase transition in their specific heat at temperatures of around $T=1$~{\rm K} whose origin is not yet understood. It is, however, well known that these single crystals suffer from impurities with $5-10$\% of the chemical composition Pb$_{2}$Te$_{3}$O$_{8}$ (Supplementary Note 3). This is in contrast to the higher quality powder samples which do not show evidence of any transition around $1$~{\rm K}, hence, supporting the view that the low temperature transition in single crystals results from disorder and strain effects. While this transition could, in principle, mark the onset of magnetic order, however our analysis (Supplementary Note 4) at least rules out the most obvious types of order which prompts us to speculate that the transition is of structural type. In this work, however, we will not focus on possible effects of impurities but rather discuss the physics of the cleaner powder samples and single crystals above the temperature/energy scale of $1$~{\rm K}.

\subsection*{Diffuse continuum of excitations}
\begin{figure}
\centering
\includegraphics[width=0.9 \textwidth]{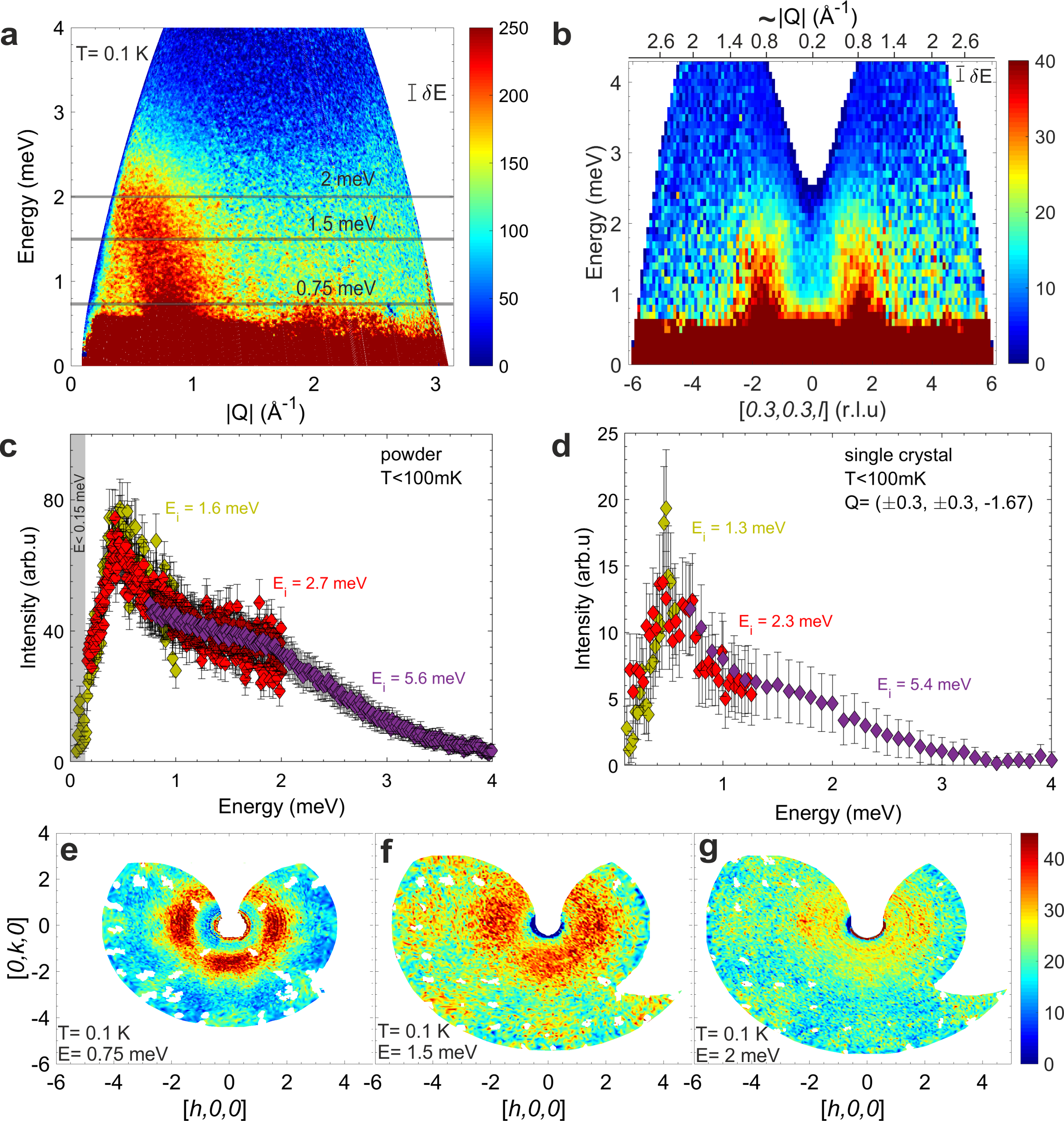}
\caption{\sffamily \textbf{Magnetic inelastic neutron scattering data of \pcto measured at temperatures of $\mathbf{T<0.1~{\rm{K}}}$}. \textbf{a} \& \textbf{b} Excitation spectra obtained on powder and single crystal using the time-of-flight spectrometer LET, with an incident energy of $E_{i}=5.6$~meV and $E_{i}=5.4$~meV respectively. These plots clearly show the presence of a diffuse, dispersionless continuum of magnetic excitations originating at $|\mathbf{Q}|\approx0.8$~\AA$^{-1}$ extending up to $E\approx{3}$~meV. \textbf{c} Powder magnetic intensity plotted as a function of energy for data collected on LET with $E_{i}=5.6$~meV (resolution $\delta{E}=0.18$~meV, integration range $0.5\leqslant|\mathbf{Q}|\leqslant1.2$ \AA$^{-1}$), $E_{i}=2.7$~meV ($\delta{E}=0.08$~meV, $0.55\leqslant|\mathbf{Q}|\leqslant1.1$ \AA$^{-1}$) and $E_{i}=1.6$~meV ($\delta{E}=0.03$~meV, $0.56\leqslant|\mathbf{Q}|\leqslant1.0$ \AA$^{-1}$). The background of each dataset has been subtracted and the datasets have been normalized to each other. The shaded area (E$<$0.15~meV) indicates the region below which data is unreliable due to subtraction of the incoherent background. The error bars here represent statistical errors. \textbf{d} Similiar plot for single crystal data measured on LET with incident energies $E_{i}=5.4$~meV (resolution $\delta{E}=0.18$~meV, integration range $0\leqslant\lbrack{h,h,0}\rbrack\leqslant0.6$~r.l.u, $-2.75\leqslant\lbrack{0,0,l}\rbrack\leqslant-0.25$~r.l.u), $E_{i}=2.3$~meV ($\delta{E}=0.06$~meV, $-1\leqslant\lbrack{h,h,0}\rbrack\leqslant1$~r.l.u, $-1.95\leqslant\lbrack{0,0,l}\rbrack\leqslant-0.95$~r.l.u) and $E_{i}=1.3$~meV ($\delta{E=0.03}$~meV, $0.95\leqslant\lbrack{h,h,0}\rbrack\leqslant2.4$~r.l.u, $-2.97\leqslant\lbrack{0,0,l}\rbrack\leqslant2.97$~r.l.u). \textbf{e-g} Single crystal spectra measured on the MACS spectrometer in the $\lbrack{h,k,0}\rbrack-$plane at constant energy transfers of E$=0.75$~meV, $1.5$~meV, and $2$~meV respectively. Non-magnetic features such as phonons and Bragg peak tails have been removed from the spectrum. The color map at $0.75$~meV reveals a broad, diffuse ring-like feature at $|\mathbf{Q}|\approx0.8$~\AA$^{-1}$ whose intensity modulates with maxima at $(1.69,\pm0.3,0)$, and equivalent positions. Similar features are present at higher energies with reduced intensity. The uncertainties in figures \textbf{c-d} represent s.e.m.}
\label{Figure:3}
\end{figure}

To explore the magnetic excitations of \pcto, we performed inelastic neutron scattering. This technique directly measures the dynamical structure factor S$(\mathbf{Q},E)$, which is the Fourier transform in space and time of the spin-spin correlation function and allows the magnetic excitation spectrum to be mapped out as a function of energy $E$ and momentum (or wavevector) transfer $\mathbf{Q}$. Figure~\ref{Figure:3}a shows the excitation spectrum of a powder sample measured at $T=0.1$~{\rm K}. A dispersionless, broad diffuse band of magnetic signal is clearly visible around momentum transfer $|\mathbf{Q}|\approx0.8$~\AA$^{-1}$. The magnetic excitations extend up to $3$ meV and are much broader than the instrumental resolution. Figure~\ref{Figure:3}c shows the magnetic signal at $|\mathbf{Q}|\approx0.8$~\AA$^{-1}$ plotted as a function of energy. The intensity is greatest at $E=0.5$~meV and weakens gradually with increasing energy. The intensity also decreases rapidly with decreasing energy and the presence of an energy gap smaller than $0.15$~meV is possible, but cannot be confirmed within the experimental uncertainty.

To obtain a more detailed picture, inelastic neutron scattering was performed on a single crystal sample which also shows the continuous magnetic excitations at $|\mathbf{Q}|\approx0.8$~\AA$^{-1}$ extending up to $3$~meV (see Fig.~\ref{Figure:3}b and d), in agreement with the powder data. Figure~\ref{Figure:3}e-g show the momentum-resolved excitations in the $\lbrack{h,k,0}\rbrack$-plane measured at the constant-energy transfers $E=0.75$, $1.5$ and $2$~meV, respectively, while Fig.~\ref{Figure:4}b gives the scattering at $E=0.5$~meV. For all energy transfers, the excitations form a diffuse ring at $|\mathbf{Q}|\approx0.8$~\AA$^{-1}$, while additional weaker branches of scattering extend outwards to higher wavevectors. At low energy transfers ($E<1$~meV) the diffuse ring has double maxima at wavevectors $(1.69,\sim\pm{0.3},0)$ and $(\sim\pm{0.3},1.69,0)$, etc. (see Fig.~\ref{Figure:4}h) while at higher energies it broadens and becomes weaker. The ring can also be observed in the $\lbrack{{h,h,l}}\rbrack$-plane where its intensity also shows a modulation (as shown in Fig.~\ref{Figure:4}a for $E=0.5$~meV). Together, these results indicate that the excitations in fact form a diffuse sphere in reciprocal space  with a radius of $|\mathbf{Q}|\approx0.8$~\AA$^{-1}$. The excitations of \pcto are clearly very different from the sharp and dispersive spin-wave excitations expected in conventional magnets with long-range magnetically ordered ground states or from the gapped and dispersive magnon excitations of dimer magnets~\cite{Diana2010,Kofu2009}. The possibility of a multimagnon continuum can also be excluded since in this case sharp excitations due to single magnons would still be expected at low energies below the continuum which are not observed (Supplementary Note 5). Additionally, the stoichiometric nature of the compound rules out disorder as the origin of the diffuse inelastic spectrum. The diffuse scattering features observed in \pcto may indicate a multi-spinon continuum of excitations as has been well documented in one-dimensional antiferromagnets formed from half-integer spin magnetic ions~\cite{lake2005,Lake2013,Mourigal2013}. They have also been observed in several two-dimensional quantum spin liquids where similar diffuse ring-like features have been found~\cite{Han2012,Balz2016}. In three dimensions, most spin liquid candidates are based on the pyrochlore structure and their scattering forms a distinctive pinch-point pattern~\cite{Henley2005}.

\subsection*{Magnetic Hamiltonian}
\begin{figure}
\centering
\includegraphics[width=0.6 \textwidth]{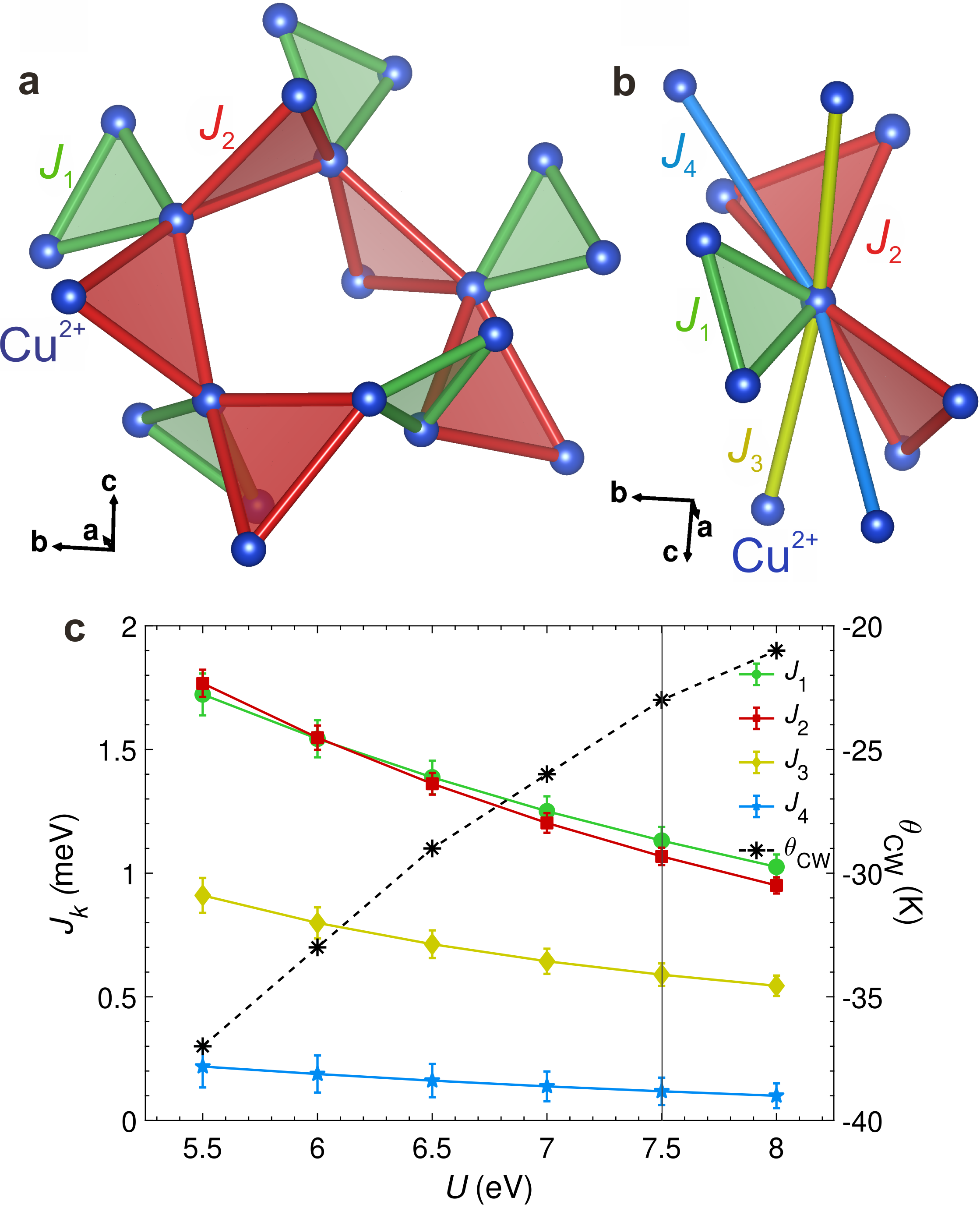}
\caption{\sffamily \textbf{The magnetic interactions and Hamiltonian of \pcto.~a} The magnetic structure drawn from the crystallographically equivalent magnetic $S=\frac{1}{2}$ Cu$^{2+}$ ions considering only the 1$^{\rm st}$ and 2$^{\rm nd}$ nearest-neighbor interactions $J_{1}$ and $J_{2}$ respectively. $J_{1}$ forms isolated equilateral triangles whereas $J_{2}$ leads to a three-dimensional network of corner-sharing triangles also known as the {\it hyperkagome} lattice. Here, the trinagles on the  outer Cu$^{2+}$ ions are dropped for simplicity. The 3$^{\rm rd}$ and 4$^{\rm th}$ neighbour couplings $J_{3}$ and $J_{4}$, respectively, are also included in \textbf{b} and couple the Cu$^{2+}$ ions into chains. The chains formed by $J_{3}$ run parallel to the cubic {\bf a}, {\bf b}, {\bf c} axes while the chains due to $J_{4}$ follow the body diagonals. \pcto crystallizes in {\it cubic} symmetry with {\it space group $P\,4_{1}32$} and the Cu$^{2+}$ ions occupy a single Wyckoff site~\cite{Wulff1997}. The graph in \textbf{c} shows the strengths of the four nearest neighbour interactions as a function of Cu$^{2+}$ onsite interaction $U$ calculated by density functional theory (coloured symbols, left hand axis). These calculations were performed with the full potential local orbital basis (FPLO) set~\cite{Koepernik1999}, and the generalized gradient approximation functional~\cite{Perdew1996}; the coupling constants were then determined by fitting to the Hamiltonian in Eq.~\ref{eq:1}. The error bars indicate statistical errors of the fit. The Curie-Weiss temperature was calculated for each set of exchange constants using $\theta_{\rm CW}=-\frac{S(S+1)}{3k_{b}}\sum_{k=1}^{4}z_{k}J_{k}$ (for single-counting of bonds) where $z_{k}$ is the coordination number of the $J_{k}^{\rm th}$ interaction (black stars, right hand axis). All the interactions are antiferromagnetic and they have the ratio $J_{1}{:}J_{2}{:}J_{3}{:}J_{4}\approx1{:}1{:}0.5{:}0.1$. $J_{1}$ and $J_{2}$ are approximately equal, and much stronger than $J_{3}$ and $J_{4}$. Together, $J_{1}$ and $J_{2}$ result in the hyper-hyperkagome lattice where each magnetic ion participates in three corner-sharing triangles forming closed loops of 4 spins and 6 spins (compared to two triangles and 10-spin loops in the hyperkagome latice).}
\label{Figure:1}
\end{figure}

Having confirmed that \pcto exhibits features characteristic of a quantum spin liquid, we now investigate the origins of this behaviour by deriving the exchange interactions. For this purpose, we employ density functional theory (DFT). The resulting values of the interaction strengths are plotted as a function of the onsite interaction $U$ in Fig.~\ref{Figure:1}c for $U=5.5$~eV to $8$~eV as this range spans the usual values for Cu$^{2+}$. We find that all the interactions are antiferromagnetic. In contrast to previous perturbation theory based DFT calculations where the hyperkagome interaction $J_{2}$ was found to be much stronger than the other interactions~\cite{Koteswararao2014}, our significantly better approach of energy mapping within DFT reveals that the two frustrated interactions $J_{1}$ and $J_{2}$ are of almost equal strength and are significantly stronger than the chain interactions $J_{3}$ and $J_{4}$. The combined effect of $J_{1}$ and $J_{2}$ is to couple the Cu$^{2+}$ ions into a highly frustrated three-dimensional network of corner-sharing triangles similar to the hyperkagome lattice ($J_{2}$ only) but with a higher density of triangles. In the hyperkagome lattice each magnetic ion participates in two corner-sharing triangles, while in \pcto each Cu$^{2+}$ ion participates in three triangles resulting in a higher connectivity - we name this lattice the hyper-hyperkagome. An important difference between these two lattices is the size of the smallest possible closed loops (beyond the triangles) around which the spins can resonate. The hyperkagome lattice consists of interconnected loops of $10$ spins. The hyper-hyperkagome can also be viewed as interconnected loops, however, with the smallest connecting $4$ spins and another consisting of 6 spins. For comparison, the 2D kagome has smallest loops of 6 spins (see Fig.~\ref{Figure:1}a). As shown in Fig.~\ref{Figure:1}c, the values of the exchange interactions decrease as the value of $U$ increases. For each value of $U$ the resulting set of interactions strengths can be used to calculate the Curie-Weiss temperature $\theta_{\rm CW}$. Since DC susceptibility measurements yield $\theta_{\rm CW}=-22$~{\rm K}~\cite{Koteswararao2014,Khuntia2016}, we use $U=7.5$~eV (corresponding to $\theta_{\rm CW}=-23$~{\rm K}) giving interaction sizes $J_{1}=1.13$~meV, $J_{2}=1.07$~meV, $J_{3}=0.59$~meV, and $J_{4}=0.12$~meV ($J_{1}{:}J_{2}{:}J_{3}{:}J_{4}\approx 1{:}1{:}0.5{:}0.1$). These values are significantly different from the reported ratio $J_{1}{:}J_{2}{:}J_{3}=0.54{:}1{:}0.77$ given in Ref.~\cite{Koteswararao2014} (Supplementary Note 6).

\subsection*{Comparison to theory}
\begin{figure}
\centering
\includegraphics[width=0.65 \textwidth]{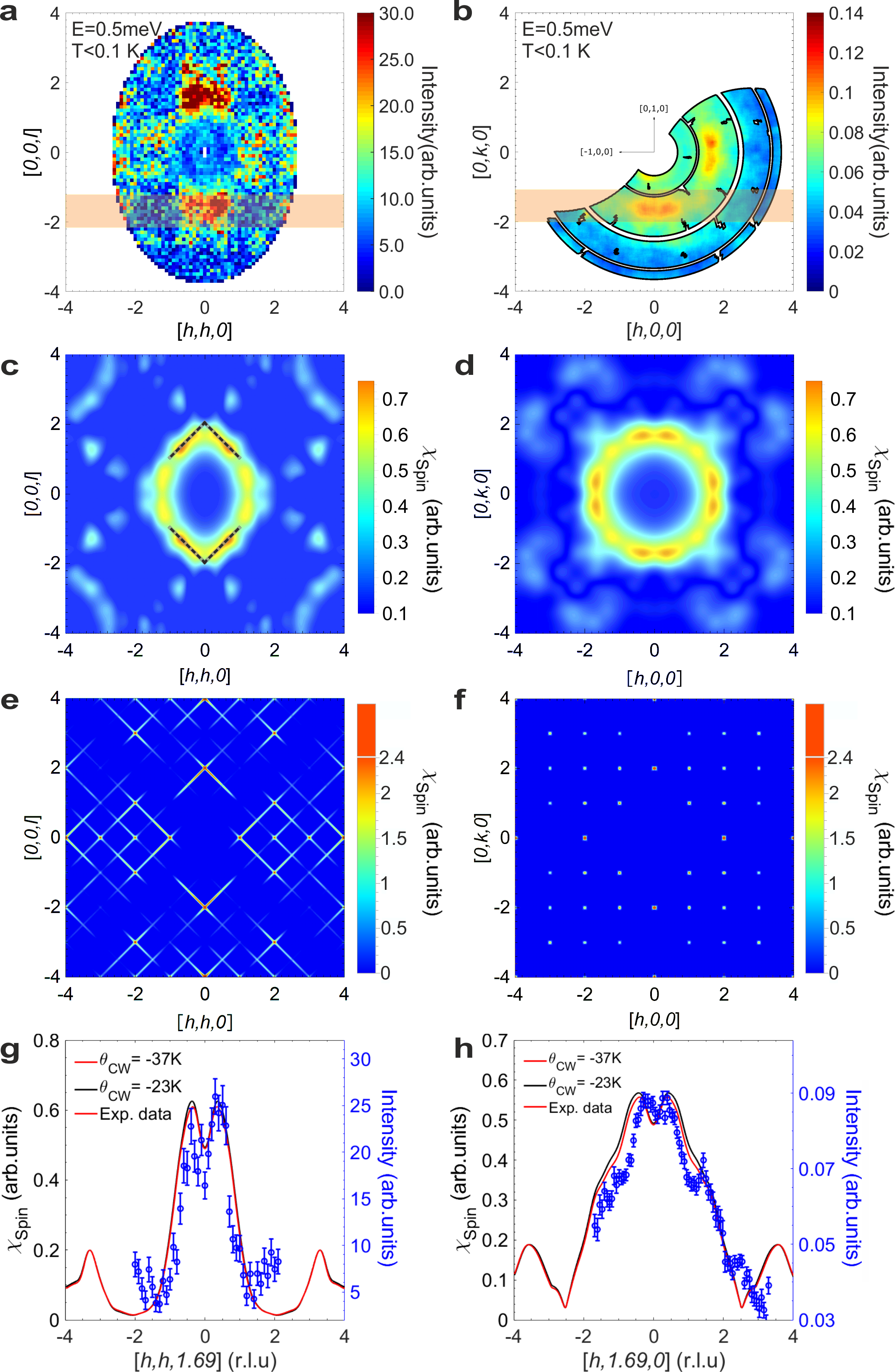}
\caption{\sffamily \textbf{The low energy magnetic excitation spectrum of \pcto compared to theory.} The color maps in \textbf{a} and \textbf{b} show the excitations measured at energy transfer $E=0.5$~meV and temperature $T<0.1$~{\rm{K}} in the $\lbrack{h,h,l}\rbrack$- and $\lbrack{h,k,0}\rbrack-$planes respectively. The $\lbrack{h,h,l}\rbrack$-map was measured using the LET spectrometer with an incident energy of E$_{i}=5.46$~meV. The data has been integrated over energy transfer $0.4\leqslant{E}\leqslant{0.6}$~meV, and out-of-plane wave-vector transfer $-0.1\leqslant\lbrack{h,-h,0}\rbrack\leqslant0.1$~r.l.u. The $\lbrack{h,k,0}\rbrack$ spectrum was measured on the ThALES spectrometer with an energy resolution of $0.097$~meV. \textbf{c} and \textbf{d} Static (real valued) spin susceptibility calculated using the pseudo-fermion functional renormalization group (PFFRG) method in the quantum limit ($S=1/2$) for $J_{1}=1.13$~meV, $J_{2}=1.07$~meV, $J_{3}=0.59$~meV, and $J_{4}=0.12$~meV (corresponding to a Curie-Weiss temperature $\theta_{\rm CW}=-23$~{\rm{K}}). The dashed black lines indicate the positions of strong scattering in the classical $J_{1}{-}J_{2}$-only model [see \textbf{e} and \textbf{f}]. \textbf{e} and \textbf{f} Corresponding classical PFFRG results obtained in the limit of large spin magnitude for a model with only $J_{1}$ and $J_{2}$ couplings of equal strength. \textbf{g} and \textbf{h} The experimental and theoretical magnetic intensity as a function of wave-vector transfer along $\lbrack{h,h,-1.69\rbrack}$ and $\lbrack{h,-1.69,0\rbrack}$ respectively. The data points (blue circles) were measured at an energy transfer $E=0.5$~meV and temperature $T{<}0.1$~{\rm{K}} and were obtained by integrating the data shown in \textbf{a} and \textbf{b} over the respective shaded regions. The solid lines show the theoretical intensity obtained by performing the same integration through the theoretical simulations shown in \textbf{c} and \textbf{d}. The theoretical intensity distribution is also shown for another set of exchange interactions represented by their corresponding Curie-Weiss temperature $\theta_{\rm CW}=-37$~{\rm{K}} (see Figure.~\ref{Figure:1}c). The calculations in \textbf{c} to \textbf{h} are corrected for the Cu$^{2+}$ form factor and the error bars in \textbf{g} and \textbf{h} represent s.e.m.}
\label{Figure:4}
\end{figure}

To gain further insight into the magnetic behaviour of \pcto, the static susceptibility expected from this set of interactions was calculated using the theoretical technique of pseudo-fermion functional renormalization group (PFFRG). This method calculates the real part of the static spin susceptibility which corresponds to the energy-integrated neutron scattering cross-section as discussed in the methods section. In agreement with the experimental observations for the powder sample, the susceptibility does not show any sign of long-range magnetic order even down to the lowest temperatures, confirming that static magnetism is suppressed by this Hamiltonian. The momentum resolved susceptibility calculated at $T=0.2$~{\rm K} is shown in Fig.~\ref{Figure:4}c-d for the $\lbrack{h,k,0}\rbrack$- and $\lbrack{h,h,l}\rbrack-$planes respectively. It predicts a diffuse sphere of scattering at the same wave-vectors and with similar intensity modulations as those observed experimentally (Fig.~\ref{Figure:3}e-g and~\ref{Figure:4}a-b), and is even able to reproduce the weaker features. The accuracy of the calculations can be further demonstrated by comparing cuts through the data and simulations. As shown in Fig.~\ref{Figure:4}g-h, the theory reproduces the double maxima as well as the structure of the slopes of these peaks to high precision. We emphasize that this level of agreement has hardly ever been achieved for such a material with many competing interactions on a complicated three-dimensional lattice and in the extreme quantum (spin-$\frac{1}{2}$) limit. From a more general viewpoint, it demonstrates that the combination of DFT and PFFRG provides a powerful and flexible numerical framework for the investigation of real quantum magnetic materials. The PFFRG method was also used to test the robustness of the spin liquid state to variations in the Hamiltonian. We find that the ground state shows no tendency toward long-range magnetic order when the ratio of interactions are varied over $0.975\leqslant{J_{1}/J_{2}}\leqslant1.08$ (corresponding to $-37$~{\rm K}$\leqslant\theta_{\rm CW}\leqslant-21$~{\rm K}) while the momentum-resolved susceptibility changes only slightly (Fig.~\ref{Figure:4}g-h).

\section*{Discussion}
In total, the neutron data and numerical simulations, together with the small spin-$\frac{1}{2}$ moments and the isotropic interactions point to the presence of strong quantum fluctuations that destroy long-range magnetic order or any static magnetism  in the ground state of \pcto. This is in stark contrast to the previously studied 3D pyrochlore classical spin ice materials with large moments and highly anisotropic interactions, where the magnetic moments are static in the ground state~\cite{fen09,mor09}. A fluctuating ground state as observed for \pcto is known to provide the right physical environment for spin fractionalization associated with deconfined spinon excitations. Such particles are generally observed as a multi-spinon spectrum that is broad and diffuse in momentum and energy. This, in turn, is the type of signal which we independently observed in both inelastic neutron experiments and PFFRG calculations making our quantum spin-liquid interpretation plausible. The issue of whether this is a gapped or gapless quantum spin liquid remains unresolved, however a clear depletion of magnetic states at low energy suggests that a gap smaller than $0.15$~meV could exist.

An important remaining question is why the complex model we propose for \pcto induces sufficiently strong quantum fluctuations for quantum spin liquid formation. According to common understanding, quantum effects for small spins are particularly strong when the corresponding classical (large spin) model exhibits an infinite ground state degeneracy, as is the case for the kagome or pyrochlore models with isotropic antiferromagnetic interactions. Performing a classical Monte Carlo analysis of our system, we found that the full $J_{1}{-}J_{2}{-}J_{3}{-}J_{4}$ model in fact does not exhibit infinite degeneracy for large spin but instead shows long-range magnetic order~\cite{Reuther2018}. However, we have identified an infinite degeneracy in the classical model with only the $J_{1}$ and $J_{2}$ interactions. From this perspective, the weaker $J_{3}$ and $J_{4}$ couplings act as perturbations inducing a small energy splitting in the degenerate classical $J_{1}{-}J_{2}$ -only system. We, therefore, propose that the strong quantum fluctuations of the full $J_{1}{-}J_{2}{-}J_{3}{-}J_{4}$ model with quantum spin-$\frac{1}{2}$ originate from the degeneracy of the classical $J_{1}{-}J_{2}$ model. This is supported by PFFRG calculations showing that the correlation profiles of both systems resemble each other (see Fig.~\ref{Figure:4}e, where the degeneracy of the $J_{1}{-}J_{2}$-only classical model manifests as streaks in the $\lbrack{1,1,1}\rbrack$ direction). Finally, the degeneracy in the classical $J_{1}{-}J_{2}$  model can be understood from the fact that the hyper-hyperkagome lattice forms a network of corner-sharing triangles. As for the classical antiferromagnetic kagome and pyrochlore lattices, the ground states in such corner-sharing geometries must obey the local constraint that the vector sum of the spins in each triangle or tetrahedron is zero. The large ground-state degeneracy then follows from the fact that there are infinitely many states which fulfill all constraints.  

In conclusion, while no experimental technique or theoretical method is able to conclusively prove the existence of a quantum spin liquid we show using a combination of theory and experiment that \pcto exhibits the measureable features expected of a quantum spin liquid including no detectable static magnetism and the presence of diffuse dispersionless spinon-like excitations. Although \pcto has a complex Hamiltonian it is clear that the frustration arises from the network of corner-sharing triangles due to the dominant $J_1$ and $J_2$ interactions. While this has been explored in the hyperkagome lattice where each spin participates in two corner-sharing triangles giving closed loops of $10$ spins~\cite{Hopkinson2007,Zhou2008,Bergholtz2010}, there has until now been little experimental or theoretical exploration of this more highly connected hyper-hyperkagome lattice where each spin participates in three corner-sharing triangles resulting in smaller closed loops of $4$ spins. The weaker interaction $J_{3}$ which reduces the cassical ground state degeneracy has the tendency to bring the quantum system closer to long-range magnetic order and may be the reason why order might be present in the single crystal samples with more impurities where impurities and additional defects could disrupt the frustration. The hyper-hyperkagome lattice has also been found in Co doped $\beta$-Mn~\cite{Paddison2013}. Here the spin is effectively classical and the metallic nature of this material promotes long-range interactions where the ferromagnetic 6$^{th}$ neighbour interaction has a similar strength to the antiferromagnetic $J_{1}$ and $J_{2}$. The wavevector-dependent scattering has some similarities to \pcto although with sharper features and can be explained by a model where $J_{6}$ ferromagnetically couples the spins into rods which then form competing triangular lattices.

In summary, three-dimensional spin liquids are very rare and current examples are confined mostly to the pyrochlore and hyperkagome lattices, thus our experimental and theoretical results are of high importance because they reveal a distinctly different type of three-dimensional lattice capable of supporting spin liquid behaviour.

\section*{Methods}
\subsection*{Neutron scattering measurements}
Powder neutron diffraction was performed on the time-of-flight diffractometer WISH at the ISIS Facility, Didcot, U.K. The sample (weight $13$~g) was placed into a copper can and the diffraction patterns were collected at $T=2$~{\rm K} and $0.1$~{\rm K}. The powder inelastic neutron scattering data was obtained at the time-of-flight spectrometer LET also located at the ISIS facility. For these measurements the same powder sample (weight $13$~g) was placed between two coaxial copper cans to achieve a cylindrical sample shape, and Helium exchange gas was used for better temperature stability. The measurements were performed at $T=0.1$~{\rm K} with incident energies: $E_i=18.2$~meV, $5.64$~meV, $2.72$~meV, $1.59$~meV. Single crystal inelastic neutron measurements in the $\lbrack{h,k,0}\rbrack-$plane were obtained at the ThALES triple-axis spectrometer using the flatcone detector at the ILL, Grenoble, France, and also at the MACS triple-axis spectrometer at NIST, Gaithersburg, USA. Wavevector maps at constant energy were measured on ThALES at $T=0.05$~{\rm K} while rotating the crystal in $0.5~\deg$ steps with a fixed final energy of $E_f=4.06$~meV giving an energy resolution $0.097$~meV. The wavevector resolution in the plots is $0.05$~r.l.u${\times}0.05$~r.l.u. At MACS, the initial energy was set to $E_i= 4$~meV for energy transfer of $E=0.75$~meV (giving energy resolution of $0.24$~meV) and  $E_i=5$~meV for $E=1.5$~meV and $2$~meV (energy resolution $0.35$~meV). The wavevector maps were obtained by rotating the crystal with a step size of $1~\deg$ and the data were plotted by rebinning to $0.04$~r.l.u${\times}0.04$~r.l.u pixels. The maps in the $\lbrack{h,h,l}\rbrack-$plane were obtained at the LET spectrometer in ISIS at $T=0.03$~{\rm K} with incident energies of $E_i=26.24$~meV, $5.46$~meV, $2.29$~meV, $1.25$~meV, and $0.79$~meV. For $E_i=5.46$~meV this gives an energy resolution of $0.18$~meV.

\subsection*{Density functional theory calculations}
We determined the parameters of the Heisenberg Hamiltonian in Eq.~\ref{eq:1} for {\pcto} using density functional theory (DFT) calculations with the all electron full potential local orbital (FPLO) basis~\cite{Koepernik1999}. We based our calculations on the structure determined via powder X-ray diffraction by Koteswararao {\it et al.}~\cite{Koteswararao2014}. The exchange couplings were extracted by mapping the total energies of many different spin configurations onto the classical energies of the Heisenberg Hamiltonian~\cite{Guterding2016}. Note that this approach is different from the second order perturbation theory estimates using $J=\frac{4t^{2}}{U}$ for the exchange interactions reported in Ref.~\cite{Koteswararao2014} which includes only the antiferromagnetic super-exchange contribution based on one virtual process. In order to increase the number of inequivalent Cu$^{2+}$ ions from one to six and thus to allow for different spin configurations, we lowered the symmetry of the crystal from $P\,4_132$ to $P\,2_1$. We converged the total energies with $6{\times} 6{\times} 6$ $\mathbf{k}$-meshes and accounted for the strong electronic correlations using a GGA+$U$ exchange correlation functional~\cite{Liechtenstein1995}. The value of the Hund's rule coupling was fixed at the typical value $J_{\rm H}=1$~eV, and the onsite correlation strength $U$ was varied between $5.5$~eV and $8$~eV. We determined the most relevant $U$ by using the constraint that the exchange couplings reproduce the experimentally determined Curie-Weiss temperature of $\theta_{\rm CW}=-22$~{\rm K}~\cite{Koteswararao2014,Khuntia2016}. This led to a DFT result for the first four exchange couplings of {\pcto} of $J_{1}=1.13$~meV, $J_{2}=1.07$~meV, $J_{3}=0.59$~meV, and $J_{4}=0.12$~meV. The full results are given in the Supplementary Note 6.

\subsection*{Pseudofermion functional renormalization group calculations}
The microscopic spin model proposed by DFT calculations is treated within the PFFRG approach~\cite{Reuther2010,Iqbal2016}, which first reformulates the original spin operators in terms of Abrikosov fermions. The resulting fermionic model is then explored within the well-developed FRG framework~\cite{Polchinski1984,Metzner2012}. Effectively, the PFFRG method amounts to generating and summing up a large number of fermionic Feynman diagrams, each representing a spin-spin interaction process that contributes to the magnetic susceptibility. In terms of the original spin degrees of freedom, this summation corresponds to a simultaneous expansion in $1/S$ and $1/N$, where $S$ is the spin magnitude and $N$ generalizes the symmetry group of the spins from SU$(2)$ to SU$(N)$. The exactness of the PFFRG in the limits $1/S\to0$ and $1/N\to 0$ ensures that magnetically ordered states (typically obtained at large $S$) and non-magnetic spin liquids (favoured at large $N$) can both be faithfully described within the same numerical framework. Particularly, due to this property, no bias towards either magnetic order or non-magnetic behaviour is built-in. In principle, the PFFRG treats an infinitely large lattice, however, spin-spin correlations are only taken into account up to a certain distance while longer range correlations are put to zero. The computation times of the PFFRG scale quadratically with the correlated volume, which in our calculations comprises $2139$ lattice sites (this corresponds to correlations up to a distance of $\approx$10 nearest-neighbour distances). Likewise, continuous frequency variables (such as the dynamics of the magnetic susceptibility) are approximated by a finite and discrete frequency grid, which leads to a quartic scaling of the computational effort in the number of grid points. In our calculations we use $64$ discrete frequencies. The central outcome of the PFFRG approach is the real part of the static and momentum-resolved magnetic susceptibility which can be directly related to the experimental neutron scattering cross section $S(\mathbf{Q},E)$ through the Kramers-Kronig relation as follows:

\begin{equation}
\chi_{real}(\mathbf{Q},0)\propto\int \frac{\chi_{img}(\mathbf{Q},E)}{E}dE
\label{eqs}
\end{equation}
where, $S(\mathbf{Q},E)\propto\chi_{img}(\mathbf{Q},E)$ at very low temperatures.

Eq.~\ref{eqs} implies that ideally PFFRG should be compared to the integral of the experimental data weighted by the inverse energy. However, it is clear from this equation that the PFFRG is dominated by the low-energy part of the neutron structure factor due to the factor of $1/E$ in the integrand. As show in Fig.~\ref{Figure:3}b, the intensity of the excitation spectrum is maximum at $E\sim0.5$~meV and decreases continuously towards smaller energy transfers. Since the excitations evolve only weakly with energy, we choose the $E=0.5$~meV dataset for comparison to the PFFRG results since it has the strongest signal.

 If a magnetic system develops magnetic order, the spin susceptibility $\chi_{real}(\mathbf{Q},0)$ manifests in a breakdown of the renormalization group flow, accompanied by distinct peaks. An important advantage of the PFFRG is that even in strongly fluctuating non-ordered magnetic phases, short-range spin correlations and their momentum profiles can be accurately calculated and compared to neutron scattering results. For consistency, the PFFRG spin susceptibility is corrected for the magnetic form factor of the Cu$^{2+}$ ion in the dipole approximation~\cite{Brown2004}. 

\subsection*{Classical simulations} 
For the numerical treatment of spin systems in the classical limit $S\to\infty$ we have employed a spin-$S$ generalization of the PFFRG approach. On a technical level, this requires the introduction of $4S$ fermionic degrees of freedom per lattice site as discussed in Ref.~\cite{Baez2017}. In the classical limit, the PFFRG equations can be solved analytically to obtain the momentum resolved magnetic susceptibility. It can be shown that the classical wave vector at which the susceptibility is strongly peaked is identical to the one predicted within the Luttinger-Tisza method~\cite{Luttinger1946}. The final susceptibility is corrected for the Cu$^{2+}$ magnetic form factor in the dipole approximation. 

\section*{Data availability}
Powder neutron diffraction data were obtained on the time-of-flight diffractometer WISH at the ISIS facility, Didcot, UK. Powder and single crystal inelastic neutron scattering data were measured on the time-of-flight spectrometer LET also at the ISIS facility. Single-crystal inelastic neutron scattering data were also collected on the triple-axis spectrometers ThALES with the flat cone option at the Institut Laue-Langevin (data available at Ref.~\cite{https://doi.org/10.5291/ill-data.4-05-662}), Grenoble, France, and MACS II at the NIST Center for Neutron Research, Gaithersburg, USA. All the raw and derived data that support the findings of this study are available from the authors upon reasonable request.

\section*{End notes}
\subsection*{Acknowledgements}
We thank K. Siemensmeyer for his help with the susceptibility measurements, D. Voneshen for his help with the inelastic neutron experiments performed on LET at the ISIS facility, Tobias M\"uller for performing classical Monte Carlo simulations and C. Baines for his help with $\mu$SR experiments on LET at SMuS facility in PSI. S.C., B.L. and A.T.M.N.I. acknowledge the support of DFG through project B06 of SFB 1143 (ID 247310070). Access to MACS was provided by the Center for High Resolution Neutron Scattering, a partnership between the National Institute of Standards and Technology and the National Science Foundation under Agreement No. DMR-1508249. Y.I. and R.T. gratefully acknowledge the Gauss Centre for Supercomputing e.V. for funding this project by providing computing time on the GCS Supercomputer SuperMUC at the Leibniz Supercomputing Centre (LRZ). J.R. is supported by the Freie Universit\"at Berlin within the Excellence Initiative of the German Research Foundation. Y.I. acknowledges the Science and Engineering Research Board (SERB), India for support through Startup Research Grant No. SRG/2019/000056. This research was supported in part by the International Centre for Theoretical Sciences (ICTS) during a visit to participate in the program The 2nd Asia Pacific Workshop on Quantum Magnetism (Code: ICTS/apfm2018/11). Y.I. acknowledges the kind hospitality of the Helmholtz-Zentrum Berlin f\''ur Materialien und Energie, Berlin, Germany where a part of the research was carried out.

\subsection*{Author contributions}
A.T.M.N.I. made the powder and single crystal samples. S.C. performed or participated in all neutron measurements, and analyzed the data with help from the other authors. J.A.R.-R., R.B., P.S. supported the INS measurements and D.K., P.M. supported the neutron diffraction measurements. B.L. participated in most measurements and directed the experimental aspects of the project. The DFT calculations were performed by H.O.J., Y.I carried out the quantum PFFRG calculations with the help of J.R. and R.T., while J.R performed the classical simulations and directed the theoretical aspects of the project. S.C. and B.L. wrote the manuscript with contributions from all authors.

\subsection*{Ethics declarations}
\noindent{\bf Competing interests:} The authors declare no competing interests.

\subsection*{Supplementary information}
Included in the attachment.

\section*{Figures}

\end{document}